\documentclass{ws-procs9x6}

\begin{document}

\title{Confinement and cut-off: a model for the pion quark distribution
function}

\author{P. Stassart, F. Bissey, J.-R. Cudell, J. Cugnon, M. Jaminon and J.-P.
Lansberg}

\address{Universit\'e de Li\`ege, D\'epartement de Physique, Institut de
Physique B.5, Sart Tilman, B-4000 LIEGE 1, Belgium}

\maketitle

\abstracts{
The pion structure function is investigated in a simple pseudo-scalar coupling
model of pion and constituent quark fields. The imaginary part of the forward
Compton scattering amplitude is evaluated. We show that the introduction of
non-perturbative effects, linked through a cut-off to the size of the pion,
allows the reproduction of important features of the pion quark distribution
function.}\par
While perturbative QCD is consistent with the $Q^2$ evolution of the structure
functions provided by deep
inelastic scattering (DIS) experiments\cite{CO98}, it is not able to
predict the
structure function at an initial value from which the $Q^2$ dependence can be
evaluated as these functions depend on non-perturbative effects, like
confinement or chiral symmetry breaking.\par
Phenomenological quark models based on chiral symmetry breaking
properties, successful in describing low-energy properties of hadrons, are expected to help us understand the connection between DIS data and
non-perturbative aspects. However, regularization procedures needed to connect
the hadron to the quark loop vary from model to model and their choice has an
impact on the structure functions that can be extracted
\cite{SH93}.\par
In this talk, we try to avoid this problem by considering a simple model of the
pion, in which the $q\overline{q}\pi$ vertex is described by a simple
pseudoscalar coupling, and where possible singularities are buried in the mass
and coupling parameters, so that the diagrams needed to calculate the pion
structure function yield a finite value for the imaginary part of the
amplitude. More details are to be found in Ref\cite{BI02}. The interaction Lagrangian reads:
\begin{equation}
{L}_{int}=ig\left(\overline{\psi}\vec{\tau}\gamma_5\psi\right).\overline{\pi
}
\end{equation}
where $g$ is the quark hadron coupling. The relevant
diagrams, up to first order in the fine structure constant $\alpha_s$ and to
second order in $g$, can be classified as "box" diagrams,where one quark connects the photon lines and "cross" diagrams,where all quark lines stand between photon and hadron vertices.The imaginary part of the amplitudes from each diagram $i$ reads
\begin{equation}
\Im T_{i\mu\nu}=Cg^2\int_{}^{}d^4kt_{i\mu\nu}D_1D_2D_3D_4
\end{equation}
where $t_{i\mu\nu}$ is the fermionic trace other the loop $i$ and $D_{1-4}$ are the
fermion propagators or the cuts on each quark line in the loop. Constant C accounts for flavour, charge and  momentum integration factors. Summing all diagrams we can then identify the structure functions $W_1$ and $W_2$  
\begin{equation}
W_{\mu\nu}=\frac{1}{2\pi}\Im
T_{\mu\nu}=\left(-g_{\mu\nu}+\frac{q_{\mu}q_{\nu}}{q^2}\right)W_1
+\left(p_{\mu}-q_{\mu}\frac{p.q}{q^2}\right)\left(p_{\nu}-q_{\nu}
\frac{p.q}{q^2}\right)W_2
\end{equation}\par
At this point, to interpret the pion as a collection of partons with a
probability distribution, we should need that the cross diagrams be
suppressed by a power of $Q^2$ compared to the box diagrams.
However this is not what we get. In the small $m_{\pi}$ large $Q^2$ limit,
these contributions are
\begin{equation}
W_1^{box}=\frac{5g^2}{24\pi^2}\ell n\left[\frac{2(1-x)\nu}{mq^2}-1\right]; 
W_1^{cross}=\frac{5g^2}{24\pi^2}
\end{equation}
A reason for this is that the pion we use has not a finite size.
Imposing it by requiring that the relative four-momentum squared of the
quarks inside the pion be limited to a maximum value $\Lambda^2$, one limits
the momentum transfer in the case of the cross diagrams, as no line joins the
photon quark vertices without going through a cut-off limited quark-pion
vertex, whereas the momentum may be transferred without such a limitation in
the box diagrams where one quark line joins the quark-photon
vertices directly. The cross diagrams contribute now as higher
twists, and the remaining box contribution can be interpreted in terms
of parton distributions.\par
We fix $g$ by imposing that there be only two constituent quarks in the pion,
i.e. $\int_{0}^{1}v(x)dx=\frac{1}{2}$ where $v(x)$ is the valence quark
distribution.\par
We can then calculate the momentum fraction carried by the quarks
\begin{equation}
2\langle
x\rangle=4\int_{0}^{1}xv(x)dx=\frac{\int_{0}^{1}F_2(x)dx}{\int_{0}^{1}F_1(x)dx}
\end{equation}
as the model yields naturally the Callan Gross relation $F_2=2xF_1$.\par
\begin{figure}[ht]
\centerline{\epsfxsize=3.0in\epsfbox{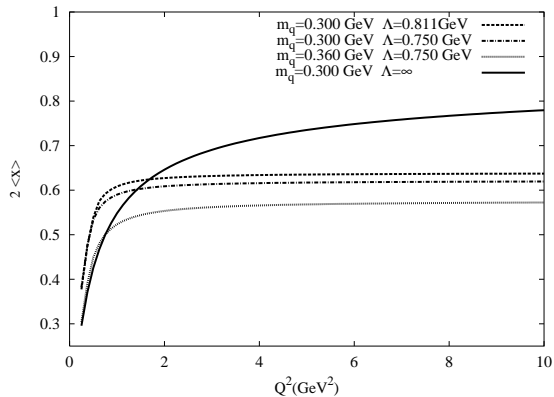}}
\caption{Momentum fraction carried by the quarks inside the neutral pion (Eq.5), as a function of $Q^2$ , for the parameter values displayed on top.}
\end{figure}
In the parton model, as $Q^2\rightarrow \infty$, $\langle 2x\rangle$ should be
equal to one. This is what we get if we do not impose that the pions have a
finite size, i.e. for $\Lambda \rightarrow \infty$.
However, when we use a physical pion, the results, shown in Fig. 1, display a
plateau at $Q^2 >2$ GeV$^2$, where the momentum fraction saturates at a value
which stays below 0.55 and 0.65 for conservative values of $m$ = 300 MeV,
$\Lambda$ = 800 MeV ; g = 3.8 is in very close agreement with the value
obtained\cite{JA92} in a NJL model for the same parameter values, which in that case
correspond to setting the correct value for $f_{\pi}$.\par
We have displayed how a cut-off is needed to represent the pion as a physical particle, from which structure functions can be deduced. When quarks behave as free particles, the sum rule holds, whereas in
the physical pion case, at least one of the quarks remains off-shell. This
implies that the quark momentum transfer is reduced by the imposition of the
cut-off at the vertex, i.e., it is suppressed by the non-perturbative
effects which the cut-off stands for. Higher twist terms
disappear then for $Q^2 >\ 2$ GeV$^2$. The quark momentum fraction is
reduced somewhat more than in the case of other hadrons. This is presumably due to
the Goldstone nature of the pion, as in the $m\rightarrow 0$ limit the momentum sum rule
is recovered.

\end{document}